\documentclass[letterpaper, 10 pt, conference]{ieeeconf}  % Comment this line out if you need a4paper

\IEEEoverridecommandlockouts                              % This command is only needed if 
\usepackage{cite}
\usepackage{amsmath,amssymb,amsfonts}
\usepackage{algorithmic}
\usepackage{graphicx}
\usepackage{textcomp}
\usepackage{xcolor}
\usepackage{array,booktabs}
\usepackage{stfloats} 
\usepackage{comment}
\usepackage{multirow}
\usepackage{float}
\def\BibTeX{{\rm B\kern-.05em{\sc i\kern-.025em b}\kern-.08em
    T\kern-.1667em\lower.7ex\hbox{E}\kern-.125emX}}
\title{\LARGE \bf
A Deep Learning approach for Depressive Symptoms assessment in Parkinson's disease patients using facial videos
}

\author{Ioannis Kyprakis$^{1}$, Vasileios Skaramagkas$^{1}$, Iro Boura$^{2}$, Georgios Karamanis$^{3}$, Dimitrios I. Fotiadis$^{4}$, \\ Zinovia Kefalopoulou$^{3}$, Cleanthe Spanaki$^{2}$, and Manolis Tsiknakis$^{1}$% <-this % stops a space
\thanks{*This work was partially supported by a grant entitled STRATIFYHF (Trustworthy artificial intelligence (AI) tools to predict the risk of chronic non-communicable diseases and/or their progression) funded by the European Commission’s HORIZON programme under Grant Agreement 101080905.}% <-this % stops a space
\thanks{$^{1}$Ioannis Kyprakis, Vasileios Skaramagkas and Manolis Tsiknakis are with the Dept. of Electrical and Computer Engineering, Hellenic Mediterranean University,  GR710 04 Heraklion, Greece and the Institute of Computer Science, Foundation for Research and Technology Hellas (FORTH), GR700 13 Heraklion, Greece 
        {\tt\small ikyprakis@ics.forth.gr}}%
\thanks{$^{2}$Iro Boura and Cleanthe Spanaki are with the School of Medicine, University of Crete, GR-710 03 Heraklion, Greece and the Dept. of Neurology, University Hospital of Heraklion, GR-715 00 Heraklion, Greece}%
\thanks{$^{3}$Georgios Karamanis and Zinovia Kefalopoulou are with the Dept. of Neurology, Patras University Hospital, 26404 Patras, Greece}%
\thanks{$^{4}$Dimitrios I. Fotiadis is with the Dept. of Materials Science and Engineering, Unit of Medical Technology and Intelligent Information  Systems, University of Ioannina, GR-451 10, Ioannina,  Greece and the Biomedical
Research Institute, Foundation for Research and Technology Hellas (FORTH), GR-451 10, Ioannina, Greece}%
}

\begin{document}

\maketitle
\thispagestyle{empty}
\pagestyle{empty}

%%%%%%%%%%%%%%%%%%%%%%%%%%%%%%%%%%%%%%%%%%%%%%%%%%%%%%%%%%%%%%%%%%%%
\begin{abstract}

Parkinson’s disease (PD) is a neurodegenerative disorder, manifesting with motor and non-motor symptoms. Depressive symptoms are prevalent in PD, affecting up to 45\% of patients. They are often underdiagnosed due to overlapping motor features, such as hypomimia. This study explores deep learning (DL) models—ViViT, Video Swin Tiny, and 3D CNN-LSTM with attention layers—to assess the presence and severity of depressive symptoms, as measured by the Geriatric Depression Scale (GDS), in PD patients through facial video analysis. A secondary analysis was conducted to evaluate the same parameters considering patients' medication states: one hour after dopaminergic medication intake (ON-medication) and 12 hours without medication (OFF-medication). Using a dataset of 1,875 videos from 178 patients, the Video Swin Tiny model achieved the highest performance, with up to 94\% accuracy and 93.7\% F1-score in binary classification (presence or absence of depressive symptoms) and 87.1\% accuracy with an 85.4\% F1-score in multiclass tasks (absence, mild, or severe depressive symptoms). 
\newline

\indent \textit{Clinical relevance}— This study underscores
the potential of DL-based spatiotemporal analysis for scalable, noninvasive diagnosis and monitoring of depressive symptoms in PD patients. These findings highlight the potential of AI-driven facial analysis as a screening tool or a supplementary aid in depressive symptom assessment, supporting clinical diagnosis and treatment decisions. Integrating such methods into routine clinical practice could improve early detection and personalized
treatment strategies, ultimately enhancing patients’ care and
quality of life.
\end{abstract}

\begin{figure*}[!b]               % * ⇒ two‑column float
  \hrule                          % horizontal rule
  \vspace{2pt}                    % a little gap
  {\footnotesize                  % switch to footnote‐sized type
    © 20XX IEEE. Personal use of this material is permitted.  
    Permission from IEEE must be obtained for all other uses, in any current 
    or future media, including reprinting/republishing this material for 
    advertising or promotional purposes, creating new collective works, for 
    resale or redistribution to servers or lists, or reuse of any copyrighted 
    component of this work in other works.
  }
\end{figure*}

%%%%%%%%%%%%%%%%%%%%%%%%%%%%%%%%%%%%%%%%%%%%%%%%%%%%%%%%%%%%%%%%%%%%%%%%%%%%%%%%
\section{INTRODUCTION}

Parkinson’s disease (PD) is a progressive neurodegenerative disorder that affects approximately 0.3\% of the global population, with its prevalence rising to over 1\% in individuals over 60 years old \cite{Tysnes2017}. The number of people affected by PD has doubled between 1990 and 2015, surpassing six million cases worldwide. This trend is expected to persist in the coming years \cite{stocchi2024parkinson,poewe2017parkinson}, posing significant challenges to health-care systems around the world, as mortality rates increase due to PD-related complications, including falls and aspiration pneumonia \cite{Ryu2023}. Despite the existing therapeutic options that alleviate motor and non-motor symptoms, such as dopamine replacement therapy and deep brain stimulation (DBS), no definite cure for PD is currently available \cite{Nemade2021}.

PD patients develop motor and non-motor symptoms, both of which significantly impact their quality of life (QoL). Key motor symptoms include bradykinesia, rigidity, tremor, hypomimia and speech difficulties \cite{Jankovic2008,Bianchini2024}, while non-motor symptoms, such as depression, cognitive decline, and sleep disturbances, can predominate in particular PD endophenotypes \cite{sauerbier2016non}, further challenging their management and reducing patients’ QoL \cite{Gonzalez-Usigli2023,Chen2020}. Addressing these symptoms requires a multi-disciplinary approach to ensure comprehensive and personalized care.

Depressive symptoms, affecting about 40-45\% of PD patients \cite{Cummings1992,Li2023}, are closely related to diminished QoL, increased disability, and faster progression of motor symptoms \cite{Khedr2020}. However, they are frequently underdiagnosed due to overlapping patterns with other PD-related symptoms, such as hypomimia, which complicate emotional assessments \cite{Goodarzi2017,Shulman2002}. This highlights the need for specialized monitoring tools to improve awareness and accurate detection.

Hypomimia–often referred to as ”facial masking”– significantly impacts PD patients’ social interactions and QoL. It is characterized by diminished facial expressiveness, which results, at least partly, from bradykinesia, and leads to decreased spontaneous facial movements and altered emotional expression \cite{Bianchini2024}. Research in facial expression analysis has shown its potential as a biomarker to understand the progression of PD and related symptoms. Specifically, Hu et al. \cite{Wei} developed a method for diagnosing PD by analyzing facial expressions through video analysis, highlighting the feasibility of facial analysis in early detection efforts. Furthermore, research into automated video-based assessments has demonstrated the ability to objectively measure facial bradykinesia in de novo PD patients, underscoring the importance of facial features in early PD diagnosis \cite{novotny2022automated}. These advancements underscore the importance of integrating facial features into clinical practice to enhance monitoring accuracy and provide insights into non-motor symptoms.

Developing non-invasive, user-friendly tools for the accurate detection of depressive symptoms holds immense potential to improve the management of PD patients. Traditional approaches, such as in-clinic assessments and self-reported questionnaires, impose significant burdens, including long administration times, high costs, and interpretation complexities, making them challenging to implement in real-world settings \cite{ebesutani2012transportable}. In contrast, video-based assessments that leverage artificial intelligence (AI) offer a scalable, remote, and less intrusive alternative to detect depressive symptoms. Such approaches facilitate early interventions and personalized treatment strategies, ultimately improving diagnosis, care, and QoL for PD patients experiencing depressive symptoms.

This work investigates utilization of advanced deep learning (DL) methodologies for the assessment of depressive symptoms in PD patients through facial video analysis. By addressing the challenges posed by hypomimia, this research establishes a state-of-the-art (SOTA) approach to detect subtle facial cues linked to the presence and the severity of depressive symptoms. Leveraging a large dataset of facial videos, the models were extensively validated, ensuring robustness and generalizability to real-world clinical settings. These efforts set a benchmark for AI-driven solutions in managing non-motor symptoms in PD, providing a tool for personalized patient care.

\section{Related Work}

In recent years, video-based systems have demonstrated high accuracy in detecting depression in the general population by analyzing facial features \cite{xie2021interpreting,parikh2024exploring,shangguan2022dual}. Expanding these approaches to selected clinical populations with increased prevalence of depressive symptoms, such as PD patients, presents unique opportunities and challenges. Researchers such as Bandini et al. \cite{bandini2017analysis} highlighted reduced facial expressivity in PD patients compared to healthy individuals, while Ge Su et al. \cite{GeSu} demonstrated that the incorporation of temporal data significantly improved the accuracy of hypomimia detection. In addition, video analysis was found effective in monitoring motor symptoms of PD, such as bradykinesia and tremor \cite{mifsud2024detecting,jin2020diagnosing,yin2021assessment}.

Despite these advances, video-based methods for identifying depressive symptoms in PD remain underexplored. Hypomimia complicates the evaluation of facial expressions to assess emotion, highlighting the need for specialized tools. The ability of DL models to analyze subtle facial movements has a significant potential to bridge this gap by improving diagnostic precision and inform treatment strategies for depressive symptoms in PD patients.

\section{Methodology}

\subsection{Dataset}

Participants were recruited from June 2023 to May 2024 from the Movement Disorders Outpatient Clinics of two Greek hospitals, the University General Hospital of Heraklion (UGHH) and the University General Hospital of Patras (UGHP). All of them met the Movement Disorder Society (MDS) criteria for Clinically Probable PD \cite{postuma2015mds}. Eligible patients were adults capable of independent ambulation or using aids. Exclusion criteria included any conditions anticipated to compromise adherence, including dementia. Patients using device-aided therapies, such as DBS or infusion pumps, were included if inclusion criteria were met. Written informed consent was obtained and the study received approval from the Ethics Committees of the two hospitals (11692/19-05-2023 and 347/13-07-2023).

The regimen of each participant was recorded. Patients on levodopa or dopamine agonists were instructed to discontinue 
their medications 12 hours prior to the initial assessment to 
achieve the OFF-medication state (often referred to as "practically defined OFF state"). Following this evaluation, 
regular medication was resumed, and a second assessment was 
conducted 60–90 minutes later in the ON-medication state. The same principles applied for DBS patients who were using levodopa and/or dopamine agonists, while they were stimulation ON under all circumstances. Drug-naïve patients, defined as those who were not treated with either levodopa or dopamine agonists, were assessed only in the OFF-medication state. Patients with advanced PD who were not able to reach the research site after a 12-hour abstinence from their medication were evaluated only in the ON-medication state. Similarly, those using a levodopa-carbidopa intestinal gel (LCIG) infusion pump were evaluated only in the ON-medication state. 

The study included 183 patients (100 from UGHH and 83 from UGHP) with PD (109 men, 74 women; mean age 65.3 ± 9.67 years). Of them, 150 completed assessments in both ON- and OFF-medication states, 18 in the ON-medication state only, and 15 in the OFF-medication state only, including 12 drug-naïve patients. Among the 11 DBS-treated patients, nine completed both assessments and two were evaluated only in the ON-medication state due to advanced PD. Detailed information on the study protocol is provided in \cite{Skaramagkas2025}.

The severity of depressive symptoms was evaluated using the Geriatric Depression Scale (GDS) \cite{gds}, a widely used and validated tool to assess depressive symptoms, including in PD patients \cite{goodarzi2016detecting}. The GDS uses thresholds to classify severity, with scores of 0-9 indicating that there are no depressive symptoms, 10–19 indicating mild and 20–30 severe depressive symptoms. For this study, 58 patients had no depressive symptoms, while 95 exhibited mild and 25 severe depressive symptoms respectively.

\begin{figure}[!htbp]
\centerline{\includegraphics[scale=0.41]{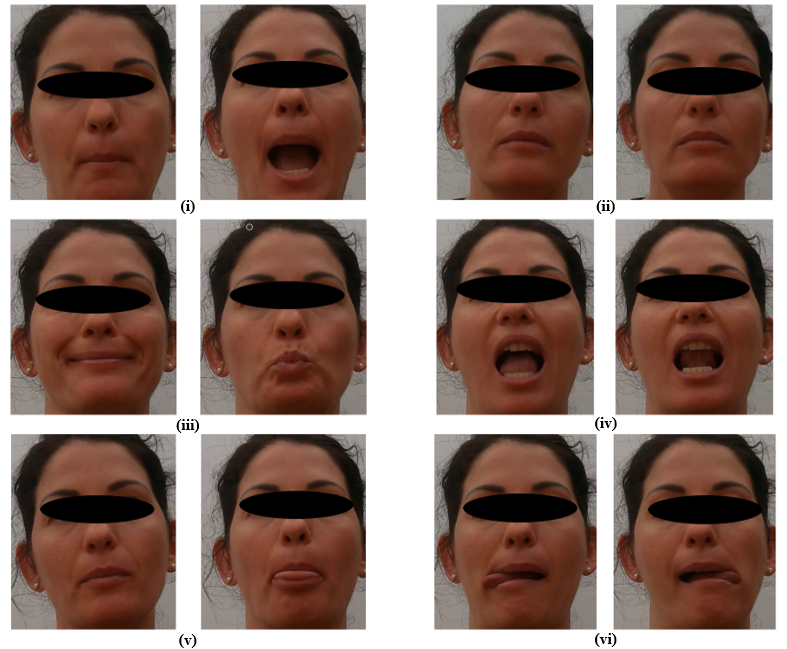}}
\caption{The six tasks performed by the PD participants; i) open and close mouth, ii) still face, iii) smile and kiss, iv) tongue up and down, v) tongue in and out, vi) tongue right and left.}
\label{figure1}
\end{figure}

The video data were recorded at 4K resolution (3840x2160 pixels) and 30 frames per second, with session lengths varying according to the assigned tasks. The recordings were taken in a controlled and quiet setting with a white backdrop and consistent illumination of approximately 450 lux. Participants performed six facial expression tasks targeting specific motor actions affected by hypomimia, shown in Fig. \ref{figure1}, repeating each task ten times as quickly as possible. The study aimed to assess both the presence and severity of depressive symptoms in Parkinson's disease (PD) patients using facial video analysis. To ensure high data quality and relevance, a rigorous selection process was applied, resulting in a final dataset of 178 patients (150 participated in the ON- and OFF-medication state, 16 in the ON-medication state and 12 in OFF-medication state). A total of 1,875 video recordings were analyzed.

\subsection{Preprocessing}

Before analyzing videos, a multistage preprocessing pipeline was implemented to ensure data homogeneity. Each step systematically standardized the input videos, improving consistency and reliability for subsequent model analyses.

\textit{1) Video Segmentation and Facial Alignment:} 
Facial alignment was performed on each frame using the Multi-task Cascaded Convolutional Neural Network (MTCNN) model \cite{mtcnn}, a DL framework designed for precise facial detection and alignment. The MTCNN three-stage cascade architecture identifies facial landmarks, detects faces, and refines the detection results, reducing inconsistencies in head posture and camera angle. The frames were then resized to \( 224 \times 224 \) pixels, allowing detailed extraction of spatial characteristics while minimizing computational load. Videos were standardized to a uniform length of 10 seconds (300 frames) to address varying durations. Longer videos were trimmed, while shorter ones were padded by repeating initial frames to avoid content loss. Additionally, all 10-second videos were segmented into 1-second clips (30 frames each) to facilitate finer-grained analysis. Pixel intensity values were normalized to a range of 0 to 1 to simplify distributions, and histogram equalization was applied to enhance contrast, making facial features more distinguishable and improving feature extraction accuracy.

\textit{2) Data augmentation:} We applied a range of data augmentation techniques to the frames. Specifically, we used small rotations to introduce minor variations in head orientation, enhancing the model's robustness to slight pose changes. Horizontal flipping increased data diversity, allowing the model to handle symmetrical variations in facial features. Moreover, Gaussian noise was added to mimic natural variations and minor imperfections in video quality, allowing the model to identify features under mildly noisy conditions. These augmentations not only helped the model learn stable representations of facial features—making it more resilient to changes in lighting, orientation, and partial obstructions—but also increased the number of training samples to meet the data requirements of our models, improving their generalization capability.

\subsection{Model Architectures}

In our investigation of depressive symptoms in PD patients, three distinct models—Video Vision Transformer (ViViT) \cite{arnab2021vivit}, Video Swin Transformer \cite{liu2022video}, and 3D-CNN-LSTM with attention layers \cite{3d-cnn-lstm}—were selected to represent a diverse set of methodologies, enabling a comprehensive assessment of their performance in detecting nuanced changes in facial behavior. ViViT leverages pure Transformers to handle long-range temporal dependencies in video tasks, the Video Swin Transformer employs hierarchical architectures with local inductive biases for efficient and precise feature extraction, and the 3D-CNN-LSTM combines convolutional layers with recurrent networks and attention mechanisms to emphasize temporal dynamics. By capturing spatiotemporal features—both spatial configurations and temporal evolutions of facial expressions—these models allow for nuanced detection of depressive symptoms. This approach enables DL models to assess depressive symptoms severity more accurately by analyzing the progression and intensity of facial expressions over time, rather than relying solely on static images.

\textit{1) ViViT:} ViViT processes video data by embedding it into a sequence of spatiotemporal tokens. The input video \( V \in \mathbb{R}^{T \times H \times W \times C} \), where \(T\), \(H\), \(W\), and \(C\) are the number of frames, height, width, and channels, respectively, is tokenized into a sequence \( z \in \mathbb{R}^{N \times d} \), where \(N\) is the number of tokens and \(d\) the token dimension. The self-attention mechanism central to ViViT computes dependencies as:
\begin{align}
\text{Attention}(Q, K, V) = \text{Softmax}\left(\frac{QK^T}{\sqrt{d_k}}\right)V,
\end{align}
where \(Q\), \(K\), and \(V\) are the query, key, and value matrices, and \(d_k\) is the dimension of the keys. ViViT reduces computational complexity by employing tubelet embedding, where the number of tokens \(N\) is:
\[
N = \left\lfloor \frac{T}{t} \right\rfloor \times \left\lfloor \frac{H}{h} \right\rfloor \times \left\lfloor \frac{W}{w} \right\rfloor,
\]
with \(t\), \(h\), and \(w\) representing tubelet dimensions. To further improve efficiency, ViViT factorizes attention into spatial and temporal components:
\begin{align}
\text{Attention}_\text{spatial} &= \text{Softmax}\left(\frac{Q_s K_s^T}{\sqrt{d_k}}\right)V_s, \\
\text{Attention}_\text{temporal} &= \text{Softmax}\left(\frac{Q_t K_t^T}{\sqrt{d_k}}\right)V_t. 
\end{align}

These operations enable ViViT to effectively capture global spatiotemporal dependencies, crucial for detecting subtle facial cues in PD patients. 

\textit{2) Video Swin Transformer:} Video Swin Transformer extends the Swin Transformer framework by employing a 3D shifted window attention mechanism for spatiotemporal data. Unlike global attention mechanisms, it computes attention locally within non-overlapping 3D patches. The self-attention is defined as:
\begin{align}
\text{Attention}(Q, K, V) = \text{Softmax}\left(\frac{QK^T}{\sqrt{d}} + B\right)V,
\end{align}
where \(B\) represents the 3D relative position bias. The shifted window mechanism introduces overlap by shifting the windows along the temporal and spatial axes:
\begin{align}
\hat{z}_l &= \text{3D-SW-MSA}(\text{LN}(z_{l-1})) + z_{l-1}, \\
z_l &= \text{FFN}(\text{LN}(\hat{z}_l)) + \hat{z}_l,
\end{align}
where \text{3D-SW-MSA} stands for \emph{3D Shifted Window Multi-Head Self-Attention}: it partitions the spatiotemporal tokens into three-dimensional windows across space and time. $\text{LN}$ and $\text{FFN}$ denote Layer Normalization and Feed-Forward Networks, respectively. In subsequent layers, these windows are shifted by a predefined amount, thereby enabling cross-window connections. Input videos of size $T \times H \times W \times C$ are divided into patches of size $P \times M \times M$, resulting in a token sequence of size:

\begin{align}
N = \left\lceil \frac{T}{P} \right\rceil \times \left\lceil \frac{H}{M} \right\rceil \times \left\lceil \frac{W}{M} \right\rceil.
\end{align}
This hierarchical structure allows the Video Swin Transformer to extract multiscale features efficiently, making it well suited to analyze localized patterns in hypomimia. For our analysis, we used the tiny version of the Video Swin transformer. The architectural design of ViViT and Video Swin Transformer tiny can be seen in Table \ref{table:table1}.

\newcommand{\mysplit}[1]{%
  \begin{tabular}{@{}c@{}}   %% removed [t]
    #1
  \end{tabular}
}

\begin{table}[!htb]
\caption{Architectural Design of transformers}
\centering
\begin{tabular}{|c|>{\centering\arraybackslash}p{0.6\linewidth}|}
\toprule
\textbf{Model} & \textbf{Parameters} \\
\midrule
\textbf{ViViT} & \mysplit{Image Patch Size: $8$\\ Frame Patch Size: $4$\\ Embedding Dimension: $128$\\ Spatial Transformer Depth: $4$\\ Temporal Transformer Depth: $4$\\ Number of Attention Heads: $4$ \\ Multilayer Perceptron Dimension: $512$} \\
\midrule
\textbf{Swin3D\_t} & \mysplit{Image Patch Size: $4$\\ Frame Patch Size: $2$\\ Embedding Dimension: $96$\\ Layer numbers: 
$\{2, 2, 4, 2\}$\\ Number of Attention Heads: $\{3, 6, 12, 24\}$ \\ Multilayer Perceptron Dimension: $384$} \\
\bottomrule
\end{tabular}
\label{table:table1}
\begin{center}
\footnotesize{Note: Parameters not mentioned are set to default values.}
\end{center}
\end{table}

\begin{figure*}[!htbp]
\centerline{\includegraphics[scale=0.56]{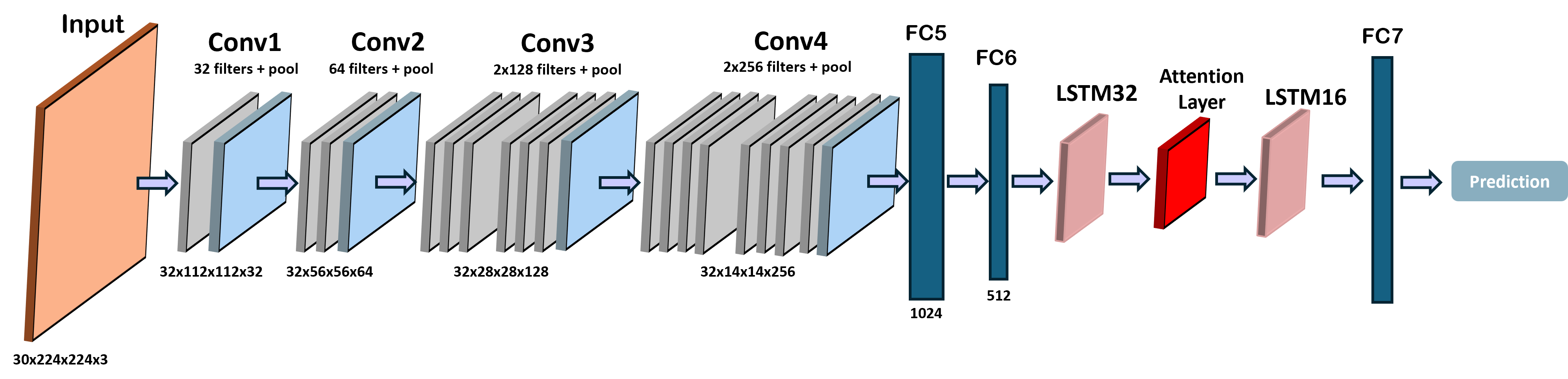}}
\caption{3D CNN-LSTM with attention architecture.}
\label{cnn_lstm}
\end{figure*}

\textit{3) 3D CNN-LSTM with Attention:} The 3D CNN-LSTM  architecture Fig. \ref{cnn_lstm} combines convolutional, recurrent, and attention-based layers to capture spatiotemporal dependencies. The 3D convolution operation extracts local features as:
\begin{align}
\text{Conv3D}(x) = \text{ReLU}(W * x + b),
\end{align}
where \(W\) is the convolution kernel, \(x\) is the input feature map, and \(b\) is the bias term. This is followed by max-pooling:
\begin{align}
\text{Pooling}(x) = \max_{p \in \text{window}}(x),
\end{align}
which reduces spatial dimensions while retaining essential features. Temporal dependencies are modeled using LSTMs. At each time step \(t\), the input is processed using gates:
\begin{align}
  &f_t = \sigma(W_f x_t + U_f h_{t-1} + b_f), \\
  &i_t = \sigma(W_i x_t + U_i h_{t-1} + b_i), \\
  &c_t = f_t \odot c_{t-1} + i_t \odot \tanh(W_c x_t + U_c h_{t-1} + b_c),
\end{align}
where \(f_t\), \(i_t\), and \(c_t\) are the forget, input, and cell states, respectively. An attention mechanism refines the LSTM output by assigning importance weights:
\begin{align}
\alpha_t = \frac{\exp(e_t)}{\sum_{k=1}^T \exp(e_k)}, \quad \text{where} \, e_t = \text{tanh}(W_h h_t + b_h).
\end{align}
The final output is generated using a fully connected layer with softmax activation:
\begin{align}
\text{Output} = \text{Softmax}(W_o h + b_o),
\end{align}
where \( W_o \) is the weight matrix, \( h \) is the input feature representation, and \( b_o \) is the bias term. 

This hybrid architecture effectively combines spatial feature extraction, sequence modeling, and attention-based refinement, making it suitable for video-based analysis of depressive symptoms in PD patients.

\section{Experiments and Results}

In this section, we present the details of the experiments conducted to evaluate the performance of our proposed models. 

\subsection{Experiments}

Two distinct experiments were carried out to address
different aspects of depressive symptoms assessment. The first experiment focused on the binary classification task of detecting whether an individual is experiencing depressive symptoms or not. The second experiment aimed not only at at predicting the depressive symptoms but also their severity levels (mild or severe). In addition, we conducted experiments to examine the impact of ON- and OFF-medication states, analyzing models' performance under these conditions to account for variations associated with medication state effects. For all models, we used the same parameters in both the ON- and OFF-medication  states to ensure a consistent evaluation framework, allowing us to observe the improvement in performance more objectively and comparably.
The motivation behind conducting these experiments was to provide a comprehensive assessment of the models’ capability to identify the presence of depressive symptoms, assess its severity, and evaluate performance in different medication states. By addressing both detection and severity prediction, while also considering ON- and OFF-medication states, our aim is to contribute valuable insights into automated depressive symptoms assessment, which can be critical in clinical settings.

To assess the robustness and generalizability of our models, we employed a Leave-One-Subject-Out (LOSO) cross-validation approach. In this method, each patient was left out in turn for testing, while the remaining patients were used for training. This approach enabled us to evaluate the performance of the model across individual subjects, ensuring a reliable generalization to unseen data. Furthermore, early stopping was implemented with a patience threshold, halting training when validation performance plateaued to prevent overfitting. Each model was trained for a maximum of 200 epochs, with validation loss monitored to determine the ideal stopping point. 

\begin{comment}
Additionally, we experimented with L2 regularization and dropout to reduce the risk of overfitting. For loss calculation, we employed binary and sparse categorical cross-entropy, suited to our classification tasks. 
\end{comment}

Finally, we tested various batch sizes to balance noisy updates from small sizes and overfitting from large ones, aiming to optimize model stability and accuracy. The aforementioned techniques ensured a thorough evaluation and fine-tuning of hyperparameters for robust learning across patients. Table \ref{table:table2} presents the hyperparameters utilized in our models, along with the corresponding total number of trainable parameters.

\begin{table}[htb]
\caption{Models Final Hyperparameters}
\centering
\begin{tabular}{cccc}
\hline
\multirow{2}{*}{\textbf{Hyper-parameters}} & \multicolumn{3}{c}{\textbf{Models}} \\
\cline{2-4}

 & \textbf{ViViT} & \textbf{Swin3D\_t} & {\textbf{3D CNN-LSTM}} \\
\hline
\textbf{Epochs} & 200 & 200 & 200 \\
\textbf{Optimizer} & Adam & Adam & AdamW \\
\textbf{Learning Rate} & \text{$1e{-4}$} & \text{$1e{-4}$} & \text{$1e{-3}$} \\
\textbf{LR Decay} & on Plateau & on Plateau & Cosine \\
\textbf{Batch Size} & 8 & 8 & 8 \\
\textbf{Loss Function} & Sparse CCE & Sparse CCE & Sparse CCE \\
\textbf{Trainable params} & 21.13 M & 28.2 M & 52.3 M \\
%\textbf{FLOPs} & 71.4 G & 43.8 G & 61 G \\
\hline
\end{tabular}
\label{table:table2}
\begin{center}
\footnotesize{Note: CCE: Categorical Cross Entropy. Binary Cross-Entropy was used as the loss function for binary classification tasks.}
\end{center}
\end{table}

\subsection{Results}

The evaluation of DL architectures across binary and multiclass classification tasks revealed notable differences in performance. The Video Swin Tiny model emerged as the top performer in binary classification, achieving an accuracy of 94\% and an F1-score of 93.7\%. Similarly, in multiclass classification, the Video Swin Tiny model delivered the best results, with an accuracy of 87.1\% and an F1-score of 85.4\%. These results underscore its strong predictive capability and ability to handle more complex tasks effectively. The complete results, including metrics for all models and scenarios, are presented in Table \ref{table:Results}.

\begin{table}[ht]
\centering
\caption{PERFORMANCE METRICS FOR BINARY AND MULTICLASS CLASSIFICATION OF DEPRESSIVE SYMPTOMS IN ON-/OFF-MEDICATION STATES}
\label{table:Results}
\begin{tabular}{|c|l|c|c|c|c|}
\hline
\textbf{Model} & \textbf{Experiment} & \textbf{Acc} & \textbf{Prec} & \textbf{Rec} & \textbf{F1} \\
\hline
\multirow{4}{*}{\mysplit{3D-CNN \\ LSTM}} 
    & Binary (OFF) & 89.4\% & 88.7\%  & {88.5\%}  & {88.6\%}  \\
    & Binary (ON) & {90.8\%}  & {90\%} & 89.8{\%}  & {89.9\%}  \\
    & Multi (OFF) & {81.3\%} & {79.5\%} & {78.9\%}  & {79.2\%}  \\
    & Multi (ON) & {82.6\%}  & {80.8\%}  & {80.3\%}  & {80.5\%}  \\
\hline
\multirow{4}{*}{\mysplit{ViViT}} 
    & Binary (OFF) & {90.6\%} & {90.0\%}  & {89.8\%} & {89.9\%}  \\
    & Binary (ON) & {92.2\%} & {91.7\%} & {91.5\%} & {91.6\%}  \\
    & Multi (OFF) & {83.1\%} & {81.8\%} & {81.2\%} & {81.5\%}  \\
    & Multi (ON) & {84.5\%}  & {83\%} & {82.5\%} & {82.7\%}  \\
\hline
\multirow{4}{*}{\mysplit{Swin3D\_t}} 
    & Binary (OFF) & \textbf{92.2\%}  & \textbf{93.5\%} & \textbf{93.3\%}  & \textbf{93.4\%}  \\
    & Binary (ON) & \textbf{94.0\%}  & \textbf{93.5\%} & \textbf{93.3\%}  & \textbf{93.4\%} \\
    & Multi (OFF) & \textbf{85.0\%} & \textbf{84.0\%} & \textbf{83.5\%} & \textbf{83.7\%}  \\
    & Multi (ON) & \textbf{87.1\%}  & \textbf{85.6\%}  & \textbf{85.2\%} & \textbf{85.4\%}  \\
\hline
\end{tabular}
\begin{center}
\footnotesize{Note: OFF: OFF-medication state; ON: ON-medication state.}
\end{center}
\end{table}

The analysis of performance in ON- and OFF-medication states further highlighted the robustness of the models. The ON-medication state generally facilitated higher accuracy and F1-scores due to reduced motor impairments and more stable facial expressions. The Video Swin Tiny model exhibited the most consistent performance across both states, with only minimal declines in accuracy and F1-score in the OFF-medication state. Although the ViViT and 3D CNN-LSTM models also performed competitively, their results were more variable in the OFF-medication state, emphasizing the importance of robust model design for real-world clinical conditions. These findings reinforce the versatility and reliability of the pretrained version of Video Swin Tiny model, establishing it as the most promising architecture for binary and multiclass depressive symptoms assessment tasks.

\section{Discussion}

Assessing depressive symptoms in PD presents significant challenges due to the intricate relationship between motor symptoms, such as hypomimia, and non-motor features of the disease. In this study, we leveraged spatiotemporal facial features to not only detect the presence of depressive symptoms but also estimate their severity using three distinct DL models. These models demonstrated the ability to analyze subtle facial expressions, providing valuable insights into the emotional state of PD patients. The prominence of facial expressions in identifying the presence of depressive symptoms and distinguishing between mild and severe depressive symptoms suggests that it can serve as a key digital biomarker for their diagnosis and monitoring.

Moreover, we investigated the impact of ON- and OFF-medication state on the models’ performance. Interestingly, better results were achieved during the ON-medication state. The reasons behind this finding remain obscure to us. Motor performance is anticipated to improve after the administration of dopaminergic medication. Consequently, this finding may reflect how the models focused more effectively on the detection and evaluation of depressive symptoms during the ON-medication state once the motor aspect of hypomimia was mitigated, at least partly. These observations showcase the effect of pharmacological interventions, which improved data consistency during evaluation, highlighting their role in optimizing depressive symptoms assessments in PD patients. By targeting tasks involving gross facial movements, such as smiling or mouth movements, the models achieved high predictive performance, whereas tasks requiring finer motor control highlighted areas for future refinement.

The main finding of this research is the demonstrated efficacy of the selected models in estimating the severity of depressive symptoms at varying levels, despite the challenges posed by hypomimia. Binary classification achieved accuracies ranging from 89.4\% to 94\%, while multiclass classification achieved accuracies between 81.3\% and 87.1\%. These results underscore the robustness of the Video Swin Tiny model, which outperformed the other approaches, in capturing both global and nuanced patterns associated with depressive symptoms in PD.

The primary motivation for this study is to enhance the clinical management and QoL for patients with PD. Traditional assessments of depressive symptoms in PD predominantly rely on self-reported questionnaires and clinical interviews—methods that are inherently subjective and can be confounded by motor impairments such as hypomimia. By introducing an objective, non-invasive, and scalable digital biomarker, this work aims to support clinicians in achieving earlier and more accurate diagnoses, thereby facilitating timely and personalized interventions. Beyond improving diagnostic accuracy, such advancements in digital health have the potential to significantly enhance patient care and overall well-being by enabling more proactive and tailored treatment approaches. Given that factors such as depression, disability, and cognitive decline are among the strongest determinants of QoL in PD patients \cite{schrag2000contributes}, integrating AI-driven assessments could help mitigate these effects through timely interventions and improved disease monitoring. This patient-centered approach aligns with broader efforts to optimize healthcare strategies and improve the daily lives of those living with Parkinson’s disease.

Few limitations must be addressed to improve clinical applicability. Variability in the severity of hypomimia among patients introduces sample heterogeneity, potentially affecting the generalizability of the model. The presence of depressive symptoms was evaluated using a single scale, which, although validated as a screening tool, may not adequately convey the multi-faceted nature and complexities of PD-related depressive symptoms. Moreover, our assessment did not account for the fluctuating aspect of depressive symptoms, which may be present in some PD patients \cite{bouranon}. More focused, currently available evaluation tools could be adopted in future analyses. Additionally, the computational demands of transformer-based architectures and hybrid frameworks may hinder their real-time deployment in resource-constrained settings. Addressing these limitations will require optimizing model efficiency and incorporating additional patient-specific factors, such as demographics and cognitive assessments, to further refine the predictions.

Future research will focus on expanding the dataset to encompass more diverse patient populations and a wider range of severity levels of PD. Incorporating real-time monitoring capabilities and expanding multimodal data sources such as thermal video recordings \cite{Gkikas} could further improve the generalizability and relevance of models in clinical practice. These advances have the potential to transform the assessment of depressive symptoms in PD, offering scalable, patient-centered solutions that support early intervention and personalized treatment strategies.

\section{Conclusions}

This study presents the first DL based approach leveraging facial video analysis to assess depressive symptoms in PD patients, addressing a critical limitation in traditional subjective assessments. By utilizing spatiotemporal facial features and SOTA models, our approach demonstrated high accuracy in detecting and estimating depression severity. The integration of an objective and scalable digital biomarker holds significant potential for enhancing clinical evaluations, enabling earlier and more accurate diagnoses, and supporting personalized interventions. Furthermore, the robustness of our method across ON- and OFF-medication states underscores its applicability in real-world clinical settings, including telemedicine and remote monitoring solutions. Future efforts will focus on improving model generalizability, incorporating multimodal data, and optimizing computational efficiency to ensure greater clinical adoption and ultimately enhance patient care and QoL in PD.


%%%%%%%%%%%%%%%%%%%%%%%%%%%%%%%%%%%%%%%%%%%%%%%%%%%%%%%%%%%%%%%%%%%%%%%%%%%%%%%%

%%%%%%%%%%%%%%%%%%%%%%%%%%%%%%%%%%%%%%%%%%%%%%%%%%%%%%%%%%%%%%%%%%%%%%%%%%%%%%%%

%%%%%%%%%%%%%%%%%%%%%%%%%%%%%%%%%%%%%%%%%%%%%%%%%%%%%%%%%%%%%%%%%%%%%%%%%%%%%%%%

\bibliographystyle{IEEEtran}
\bibliography{refs}

\begin{thebibliography}{10}
\providecommand{\url}[1]{#1}
\csname url@samestyle\endcsname
\providecommand{\newblock}{\relax}
\providecommand{\bibinfo}[2]{#2}
\providecommand{\BIBentrySTDinterwordspacing}{\spaceskip=0pt\relax}
\providecommand{\BIBentryALTinterwordstretchfactor}{4}
\providecommand{\BIBentryALTinterwordspacing}{\spaceskip=\fontdimen2\font plus
\BIBentryALTinterwordstretchfactor\fontdimen3\font minus
  \fontdimen4\font\relax}
\providecommand{\BIBforeignlanguage}[2]{{%
\expandafter\ifx\csname l@#1\endcsname\relax
\typeout{** WARNING: IEEEtran.bst: No hyphenation pattern has been}%
\typeout{** loaded for the language `#1'. Using the pattern for}%
\typeout{** the default language instead.}%
\else
\language=\csname l@#1\endcsname
\fi
#2}}
\providecommand{\BIBdecl}{\relax}
\BIBdecl

\bibitem{Tysnes2017}
\BIBentryALTinterwordspacing
O.-B. Tysnes and A.~Storstein, ``Epidemiology of parkinson's disease,''
  \emph{Journal of Neural Transmission (Vienna)}, vol. 124, no.~8, pp.
  901--905, August 2017. [Online]. Available:
  \url{https://doi.org/10.1007/s00702-017-1686-y}
\BIBentrySTDinterwordspacing

\bibitem{stocchi2024parkinson}
F.~Stocchi, D.~Bravi, A.~Emmi, and A.~Antonini, ``Parkinson disease therapy:
  current strategies and future research priorities,'' \emph{Nature Reviews
  Neurology}, pp. 1--13, 2024.

\bibitem{poewe2017parkinson}
W.~Poewe, K.~Seppi, C.~M. Tanner, G.~M. Halliday, P.~Brundin, J.~Volkmann,
  A.-E. Schrag, and A.~E. Lang, ``Parkinson disease,'' \emph{Nature reviews
  Disease primers}, vol.~3, no.~1, pp. 1--21, 2017.

\bibitem{Ryu2023}
D.-W. Ryu, K.~Han, and A.-H. Cho, ``Mortality and causes of death in patients
  with parkinson's disease: a nationwide population-based cohort study,''
  \emph{Frontiers in Neurology}, vol.~14, p. 1236296, 2023.

\bibitem{Nemade2021}
D.~Nemade, T.~Subramanian, and V.~Shivkumar, ``An update on medical and
  surgical treatments of parkinson’s disease,'' \emph{Aging and disease},
  vol.~12, no.~4, p. 1021, 2021.

\bibitem{Jankovic2008}
J.~Jankovic, ``Parkinson's disease: Clinical features and diagnosis,''
  \emph{Journal of Neurology, Neurosurgery, and Psychiatry}, vol.~79, no.~4,
  pp. 368--376, 2008.

\bibitem{Bianchini2024}
E.~Bianchini, D.~Rinaldi, M.~Alborghetti, M.~Simonelli, F.~D’Audino,
  C.~Onelli, E.~Pegolo, and F.~E. Pontieri, ``The story behind the mask: A
  narrative review on hypomimia in parkinson’s disease,'' \emph{Brain
  Sciences}, vol.~14, no.~1, p. 109, 2024.

\bibitem{sauerbier2016non}
A.~Sauerbier, P.~Jenner, A.~Todorova, and K.~R. Chaudhuri, ``Non motor subtypes
  and parkinson's disease,'' \emph{Parkinsonism \& related disorders}, vol.~22,
  pp. S41--S46, 2016.

\bibitem{Gonzalez-Usigli2023}
H.~A. Gonz{\'a}lez-Usigli \emph{et~al.}, ``Neurocognitive psychiatric and
  neuropsychological alterations in parkinson’s disease: a basic and clinical
  approach,'' \emph{Brain Sciences}, vol.~13, no.~3, p. 508, 2023.

\bibitem{Chen2020}
Z.~Chen, G.~Li, and J.~Liu, ``Autonomic dysfunction in parkinson’s disease:
  Implications for pathophysiology, diagnosis, and treatment,''
  \emph{Neurobiology of disease}, vol. 134, p. 104700, 2020.

\bibitem{Cummings1992}
\BIBentryALTinterwordspacing
J.~L. Cummings, ``Depression and parkinson's disease: A review,'' \emph{The
  American Journal of Psychiatry}, vol. 149, no.~4, pp. 443--454, 1992.
  [Online]. Available: \url{https://doi.org/10.1176/ajp.149.4.443}
\BIBentrySTDinterwordspacing

\bibitem{Li2023}
\BIBentryALTinterwordspacing
L.~Li, Z.~Wang, Z.~You, and J.~Huang, ``Prevalence and influencing factors of
  depression in patients with parkinson's disease,'' \emph{Alpha Psychiatry},
  vol.~24, no.~6, pp. 234--238, 2023. [Online]. Available:
  \url{https://doi.org/10.5152/alphapsychiatry.2023.231253}
\BIBentrySTDinterwordspacing

\bibitem{Khedr2020}
E.~M. Khedr, A.~A. Abdelrahman, Y.~Elserogy, A.~F. Zaki, and A.~Gamea,
  ``Depression and anxiety among patients with parkinson’s disease:
  frequency, risk factors, and impact on quality of life,'' \emph{The Egyptian
  Journal of Neurology, Psychiatry and Neurosurgery}, vol.~56, pp. 1--9, 2020.

\bibitem{Goodarzi2017}
\BIBentryALTinterwordspacing
Z.~Goodarzi and Z.~Ismail, ``A practical approach to detection and treatment of
  depression in parkinson disease and dementia,'' \emph{Neurology Clinical
  Practice}, vol.~7, no.~2, pp. 128--140, 2017. [Online]. Available:
  \url{https://www.neurology.org/doi/abs/10.1212/CPJ.0000000000000351}
\BIBentrySTDinterwordspacing

\bibitem{Shulman2002}
\BIBentryALTinterwordspacing
L.~M. Shulman, R.~L. Taback, A.~A. Rabinstein, and W.~J. Weiner,
  ``Non-recognition of depression and other non-motor symptoms in parkinson's
  disease,'' \emph{Parkinsonism \& Related Disorders}, vol.~8, no.~3, pp.
  193--197, 2002. [Online]. Available:
  \url{https://doi.org/10.1016/s1353-8020(01)00015-3}
\BIBentrySTDinterwordspacing

\bibitem{Wei}
C.~Hu, P.~Zhang, and W.~Huang, ``A novel face-based approach for the early
  diagnosis of parkinson’s disease,'' in \emph{2021 International Conference
  on Information Technology and Biomedical Engineering (ICITBE)}, 2021, pp.
  248--252.

\bibitem{novotny2022automated}
M.~Novotny, T.~Tykalova, H.~Ruzickova, E.~Ruzicka, P.~Dusek, and J.~Rusz,
  ``Automated video-based assessment of facial bradykinesia in de-novo
  parkinson’s disease,'' \emph{NPJ digital medicine}, vol.~5, no.~1, p.~98,
  2022.

\bibitem{ebesutani2012transportable}
C.~Ebesutani, A.~Bernstein, B.~F. Chorpita, and J.~R. Weisz, ``A transportable
  assessment protocol for prescribing youth psychosocial treatments in
  real-world settings: reducing assessment burden via self-report scales.''
  \emph{Psychological Assessment}, vol.~24, no.~1, p. 141, 2012.

\bibitem{xie2021interpreting}
W.~Xie, L.~Liang, Y.~Lu, C.~Wang, J.~Shen, H.~Luo, and X.~Liu, ``Interpreting
  depression from question-wise long-term video recording of sds evaluation,''
  \emph{IEEE Journal of Biomedical and Health Informatics}, vol.~26, no.~2, pp.
  865--875, 2021.

\bibitem{parikh2024exploring}
A.~Parikh, M.~Sadeghi, and B.~Eskofier, ``Exploring facial biomarkers for
  depression through temporal analysis of action units,'' \emph{arXiv preprint
  arXiv:2407.13753}, 2024.

\bibitem{shangguan2022dual}
Z.~Shangguan, Z.~Liu, G.~Li, Q.~Chen, Z.~Ding, and B.~Hu, ``Dual-stream
  multiple instance learning for depression detection with facial expression
  videos,'' \emph{IEEE Transactions on Neural Systems and Rehabilitation
  Engineering}, vol.~31, pp. 554--563, 2022.

\bibitem{bandini2017analysis}
A.~Bandini, S.~Orlandi, H.~J. Escalante, F.~Giovannelli, M.~Cincotta, C.~A.
  Reyes-Garcia, P.~Vanni, G.~Zaccara, and C.~Manfredi, ``Analysis of facial
  expressions in parkinson's disease through video-based automatic methods,''
  \emph{Journal of neuroscience methods}, vol. 281, pp. 7--20, 2017.

\bibitem{GeSu}
\BIBentryALTinterwordspacing
G.~Su, B.~Lin, J.~Yin, W.~Luo, R.~Xu, J.~Xu, and K.~Dong, ``Detection of
  hypomimia in patients with parkinson’s disease via smile videos,''
  \emph{Annals of Translational Medicine}, vol.~9, no.~16, 2021. [Online].
  Available: \url{https://atm.amegroups.org/article/view/77538}
\BIBentrySTDinterwordspacing

\bibitem{mifsud2024detecting}
J.~Mifsud, K.~R. Embry, R.~Macaluso, L.~Lonini, R.~J. Cotton, T.~Simuni, and
  A.~Jayaraman, ``Detecting the symptoms of parkinson’s disease with
  non-standard video,'' \emph{Journal of neuroengineering and rehabilitation},
  vol.~21, no.~1, p.~72, 2024.

\bibitem{jin2020diagnosing}
B.~Jin, Y.~Qu, L.~Zhang, and Z.~Gao, ``Diagnosing parkinson disease through
  facial expression recognition: video analysis,'' \emph{Journal of medical
  Internet research}, vol.~22, no.~7, p. e18697, 2020.

\bibitem{yin2021assessment}
Z.~Yin, V.~J. Geraedts, Z.~Wang, M.~F. Contarino, H.~Dibeklioglu, and
  J.~Van~Gemert, ``Assessment of parkinson’s disease severity from videos
  using deep architectures,'' \emph{IEEE Journal of Biomedical and Health
  Informatics}, vol.~26, no.~3, pp. 1164--1176, 2021.

\bibitem{postuma2015mds}
R.~B. Postuma \emph{et~al.}, ``Mds clinical diagnostic criteria for parkinson's
  disease,'' \emph{Movement disorders}, vol.~30, no.~12, pp. 1591--1601, 2015.

\bibitem{Skaramagkas2025}
V.~Skaramagkas, I.~Boura, G.~Karamanis, I.~Kyprakis, D.~I. Fotiadis,
  Z.~Kefalopoulou, C.~Spanaki, and M.~Tsiknakis, ``Dual stream transformer for
  medication state classification in parkinson’s disease patients using
  facial videos,'' \emph{npj Digit. Med.}, vol.~8, no.~1, p. 226, Apr. 2025.

\bibitem{gds}
J.~A. Yesavage, T.~L. Brink, T.~L. Rose, O.~Lum, V.~Huang, M.~Adey, and V.~O.
  Leirer, ``Development and validation of a geriatric depression screening
  scale: a preliminary report,'' \emph{Journal of psychiatric research},
  vol.~17, no.~1, pp. 37--49, 1982.

\bibitem{goodarzi2016detecting}
Z.~Goodarzi, K.~J. Mrklas, D.~J. Roberts, N.~Jette, T.~Pringsheim, and
  J.~Holroyd-Leduc, ``Detecting depression in parkinson disease: a systematic
  review and meta-analysis,'' \emph{Neurology}, vol.~87, no.~4, pp. 426--437,
  2016.

\bibitem{mtcnn}
K.~Zhang, Z.~Zhang, Z.~Li, and Y.~Qiao, ``Joint face detection and alignment
  using multitask cascaded convolutional networks,'' \emph{IEEE signal
  processing letters}, vol.~23, no.~10, pp. 1499--1503, 2016.

\bibitem{arnab2021vivit}
A.~Arnab, M.~Dehghani, G.~Heigold, C.~Sun, M.~Lu{\v{c}}i{\'c}, and C.~Schmid,
  ``Vivit: A video vision transformer,'' in \emph{Proceedings of the IEEE/CVF
  international conference on computer vision}, 2021, pp. 6836--6846.

\bibitem{liu2022video}
Z.~Liu, J.~Ning, Y.~Cao, Y.~Wei, Z.~Zhang, S.~Lin, and H.~Hu, ``Video swin
  transformer,'' in \emph{Proceedings of the IEEE/CVF conference on computer
  vision and pattern recognition}, 2022, pp. 3202--3211.

\bibitem{3d-cnn-lstm}
J.~You and J.~Korhonen, ``Deep neural networks for no-reference video quality
  assessment,'' in \emph{2019 IEEE International Conference on Image Processing
  (ICIP)}, 2019, pp. 2349--2353.

\bibitem{schrag2000contributes}
A.~Schrag, M.~Jahanshahi, and N.~Quinn, ``What contributes to quality of life
  in patients with parkinson's disease?'' \emph{Journal of Neurology,
  Neurosurgery \& Psychiatry}, vol.~69, no.~3, pp. 308--312, 2000.

\bibitem{bouranon}
I.~Boura, K.~Poplawska-Domaszewicz, C.~Spanaki, R.~Chen, D.~Urso, R.~van
  Coller, A.~Storch, and K.~R. Chaudhuri, ``Non-motor fluctuations in
  parkinson's disease: underdiagnosed, yet important,'' \emph{Journal of
  movement disorders}, 2024.

\bibitem{Gkikas}
S.~Gkikas and M.~Tsiknakis, ``Synthetic thermal and rgb videos for automatic
  pain assessment utilizing a vision-mlp architecture,'' in \emph{2024 12th
  International Conference on Affective Computing and Intelligent Interaction
  Workshops and Demos (ACIIW)}, 2024, pp. 4--12.

\end{thebibliography}

\end{document}